# Wide-band capacitance measurement on a semiconductor double quantum dot for studying tunneling dynamics

Takeshi Ota, Toshiaki Hayashi, and Koji Muraki
*NTT Basic Research Laboratories, NTT Corporation, 3-1, Morinosato-Wakamiya, Atsugi, 243-0198*

Toshimasa Fujisawa
*NTT Basic Research Laboratories, NTT Corporation, 3-1, Morinosato-Wakamiya, Atsugi, 243-0198, Japan; Research Center for Low Temperature Physics, Tokyo Institute of Technology, 2-12-1 Ookayama, Meguro-ku, Tokyo, Japan*

We propose and demonstrate wide-band capacitance measurements on a semiconductor double-quantum dot (DQD) to study tunneling dynamics. By applying phase-tunable high-frequency signals independently to the DQD and a nearby quantum-point-contact charge detector, we perform on-chip lock-in detection of the capacitance associated with the single-electron motion over a wide frequency range from hertz to a few ten gigahertz. Analyzing the phase and the frequency dependence of the signal allows us to extract the characteristic tunneling rates. We show that, by applying this technique to the interdot tunnel coupling regime, quantum capacitance reflecting the strength of the quantum-mechanical coupling can be measured.

Detection of individual charges in semiconductor quantum dots using a quantum point contact (QPC) as a sensitive mesoscopic sensor is now one of the key technologies in the field of quantum information processing with quantum dots. As first demonstrated in Ref. [1], the capacitive coupling between the QPC and the dots allows one to monitor the charge state of the dots by measuring the current flowing through the QPC. In addition to static charge detection, charge detection in the time domain is recently attracting strong interest. For example, real-time observation of single-electron motion, or single-electron counting, has also been demonstrated using a QPC charge detector [2]. Single-shot readout of a quantum bit (qubit), which is represented by a charge or spin in quantum dots, is also a recent attractive topic [3].

An alternative detection scheme is capacitance measurement. The capacitance of a quantum dot device can be described by a circuit consisting of tunnel capacitances and tunnel resistances [4], and thus the dot capacitance is sensitive to the single-electron transport [5]. Moreover, capacitance should be corrected by quantum mechanical tunneling [6] and many body corrections [7]. For a strongly coupled double quantum dot (DQD), the quantum capacitance is defined as the second derivative of the energy $E$ with respect to the gate voltage $V_G$, i.e., $C_Q \equiv d^2E/dV_G^2$ [6]. Therefore, capacitance measurement can be used to distinguish bonding and antibonding states of single-electron states or the singlet and triplet of two-electron states. In order to develop a readout device for qubits, the capacitance has to be measured at high frequency.

In this Letter, we propose and demonstrate a new experimental technique to measure impedance (both resistance and capacitance) of semiconductor quantum dots in order to study tunneling dynamics. In contrast to the radio-frequency single-electron transistor (rf-SET) based technique with a resonator [8], we developed an on-chip lock-in technique for detecting the capacitance component by applying two rf signals to the DQD and the QPC. This method eliminates the need for an auxiliary tank circuit, allowing us to study the tunneling dynamics over a wide frequency range from hertz to a few ten gigahertz. By analyzing the frequency dependence of the signal amplitude and phase, we are able to determine the tunneling rates. By applying this technique to the interdot tunneling regime, we observe capacitance signal strongly dependent on the interdot tunnel coupling, demonstrating the detection of quantum capacitance.

Figure 1(a) schematically shows the device structure and the experimental setup. A DQD is formed in a two-dimensional electron gas at the interface of a GaAs/AlGaAs heterojunction by applying negative voltages to surface Schottky metal gates. A DQD and a QPC are formed in the upper and lower channels, respectively, which are electrically isolated from each other by a gate $V_{iso}$ between them. We use the gate voltages $V_{UL}$ and $V_{UR}$ to vary the electron number in the DQD and $V_{UC}$ to tune the strength of the interdot tunnel coupling. A QPC is defined by using $V_{DR}$ (with all other gates in the lower channel grounded) and is adjusted in the tunneling regime with the conductance at $\sim e^2/h$. In the capacitance measurements, we measure temporally averaged current $\langle I_{QPC} \rangle$ flowing through the QPC. The rf signals $V_{DQD}(t)$ and $V_{QPC}(t)$ with an operating frequency $f_{op}$ and a relative phase difference $\theta$ are applied to the DQD and the QPC, respectively. Unless otherwise specified, the phase is fixed at $\theta = 0$. While some measurements were carried out with sinusoidal waves, square waves were used to highlight the non-equilibrium effects. Here, we introduce the principle of the capacitance measurement, focusing on the tunneling across the right barrier to neglect the quantum capacitance as schematically shown in Fig. 1(b). The modulation of the right dot potential with a square wave $V_{DQD}(t)$ induces a single-electron tunneling off (the upper panel) and in (the lower panel) the dot across the barrier, resulting in charge modulation in the right dot. The tunneling events are stochastically delayed with the inverse of the tunneling rate $\Gamma^{-1}$. The ensemble response $<\Delta Q_{QDR}(t)>$ therefore has a finite rise time as shown in the lower panel of Fig. 1(c). The amplitude and phase of the signal depend on the ratio between the repetition time $T_{rep}(=1/f_{op})$ and $\Gamma^{-1}$. Since the QPC is capacitively coupled to the dot, $\Delta Q_{QDR}(t)$ modulates the conductance of the QPC, i.e., $G_{QPC}(t) \propto \Delta Q_{QDR}(t)$. By applying another rf signal $V_{QPC}(t)$ to the source electrode of the QPC, we obtain averaged current $<I_{QPC}>$ expressed as $<I_{QPC}> = <G_{QPC}(t)V_{QPC}(t)> \propto <\Delta Q_{QDR}(t)V_{DQD}(t)>$. The dc current is proportional to the capacitance (conductance) when the phase difference $\theta$ is set at 0 ($\pi/2$). Such on-chip lock-in detection should work for a wide frequency range as demonstrated below.

The experimental results for the dot-lead tunneling are shown in Figs. 1(d) and (e). Here, $V_{UR}$ is swept to change the electron number $N_R$ in the right dot while keeping that in the left dot constant. Figure 1(d) shows the conventional charge detection, where $I_{QPC}$ measured with a dc bias $V_{sd-QPC} = 0.7$ mV is plotted as a function of $V_{UR}$. As in previous reports [1], $I_{QPC}$ jumps each time $N_R$ changes by one. The result of a capacitance measurement, which was carried out separately with ac voltages ($V_{DQD} = 0.34$ mV and $V_{QPC} = 0.56$ mV) and $V_{sd-QPC} = 0$ V, is shown in Fig. 1(e), where $<I_{QPC}>$ taken at $f_{op} = 1$ kHz is plotted. Peaks (labeled I to III) appear in $<I_{QPC}>$ at voltages corresponding to the jumps in $I_{QPC}$, indicating that they are associated with the charge boundaries of the DQD. These peaks are superimposed on a constant background $\delta I_{CB}$. As we show later, this constant background originates from the direct capacitive coupling between the QPC and the drain electrode to which the square waves are applied. Note that the height of the peak $\delta I_{peak}$ is almost constant (except for the one marked with an asterisk [10]), which is reasonable because the spatial configuration of the dot and hence the geometrical capacitance are almost independent of $V_{UR}$.

Figure 2(a) shows the evolution of the $<I_{QPC}>$ trace with the frequency $f_{op}$. As $f_{op}$ is increased, peaks I, II, and III become smaller. Figure 2(b) shows $\delta I_{peak}$ of peaks I to IV as a function of $f_{op}$, where we plot the value of $\delta I_{peak}$. When $f_{op}$ exceeds the tunneling rate $\Gamma$, the probability of an electron tunneling within the given period of time $1/f_{op}$ reduces and so the signal $\delta I_{peak}$ decreases as fitted by the solid lines in Fig. 2(b) [9]. From the fitting, $\Gamma^{-1}$ of peak I, II, and III is estimated to be 40, 3, and 0.4 nsec, respectively. On the other hand, $\delta I_{peak}$ of peak IV remains almost constant up to the highest frequency in Fig. 2(b), indicating that $\Gamma^{-1}$ is much higher. In this way, the on-chip lock-in measurement is successful for a wide frequency range up to 15 GHz, which can be extended by improving the low-temperature coaxial cables.

The correlation between the charge response and the reference signal can be elaborated by analyzing the signal at various phases. Figures 2(c) and (d) compare the phase evolution of the background $\delta I_{CB}$ and the peak height $\delta I_{peak}$ measured at different $f_{op}$. The background has a triangular dependence on $\theta$, which is the correlation function of two identical square waves. This means that the response of $\delta I_{CB}$ is instantaneous, which confirms that it originates from the direct capacitive coupling between the QPC and the electrode for $V_{DQD}$. In contrast, the peak signal [Fig. 2(d)] is significantly distorted with increasing $f_{op}$. The $\delta I_{peak}$ shows not only a decrease in peak amplitude but also a phase delay (indicated by arrows in the figure). As already discussed, the decrease in peak amplitude and the phase delay comes from the ratio between $T_{rep}$ and $\Gamma^{-1}$ as illustrated in Fig. 1(c). The solid lines in Fig. 2(d) indicate the $<I_{QPC}>$ calculated for each $f_{op}$ using the same time constant, $\Gamma^{-1} = 40$ nsec, obtained above by fitting the frequency-dependent data at $\theta = 0$ [Fig. 2(b)]. The good agreement attests to the consistency and validity of our analysis.

The capacitance measurement can be used to identify the excited states of a quantum dot. The peaks in Figs. 1(e) and 2(a) are broad with a flat top. The peak width is proportional to $V_{DQD}$, and some excited states of the dot may contribute a tunneling event with different $\Gamma$. Actually, some peaks show step-like features at the intermediate frequencies as indicated by arrows in Fig. 2(a). By using a similar frequency and phase analysis, we can extract individual

Γ values for the GS and ES. Here, we note that the estimated $\Gamma^{-1}$ in Fig. 2(b) corresponds to that of the GS of each peak.

Now we investigate the capacitance for the interdot tunneling. As shown in Fig. 3(a), when $V_{UL}$ ands $V_{UR}$ are varied with $V_{UC}$ = -0.2 V, the capacitance signal displays a stability diagram which is formed by a honeycomb-shaped structure characteristic of a DQD in the weak-coupling regime [4]. The honeycomb structure consists of three types of charge boundaries between the charge states $(N_L, N_R) \leftrightarrow (N_L, N_R \pm 1)$, $(N_L, N_R) \leftrightarrow (N_L \pm 1, N_R)$, and $(N_L, N_R) \leftrightarrow (N_L \mp 1, N_R \pm 1)$ which are marked 'A', 'B', and 'C', respectively, in the figure. The data demonstrate that capacitance signals appear not only for dot-lead tunneling but also for interdot tunneling (C). A closer examination of the data further reveal that the signals associated with right-dot-right-lead tunneling (A) appear as peaks, whereas those associated with left-dot-left-lead tunneling (B) and interdot tunneling (C) appear as dips. Note that the polarity of the capacitance signals reflects the relative location of the QPC and the dot. That is, the capacitance signal becomes positive (negative) if the conductance of the QPC is enhanced (reduced) by the tunneling. Notably, in the region where both $V_{UR}$ and $V_{UL}$ are largely negative (encircled by the dashed line in the figure), only signals due to interdot tunneling are visible, with all other lines defining the honeycomb structure almost vanishing. This happens because in this region the tunnel coupling with leads is so weak that dot-lead tunneling hardly takes place within the given period of time, $1/f_{op}$. It is also worth emphasizing that the interdot tunneling is detectable even though the dots are well-isolated from the leads and no charge transfer on or off the DQD takes place.

Figure 3(c) shows the variation of the capacitance dip due to interdot tunneling with $V_{UC}$. The data are plotted as a function of the bias offset ε between the dots. (The sweep direction of ε is shown in Fig. 3(a).) As the interdot coupling is strengthened by making $V_{UC}$ less negative, the dip becomes broader and smaller. At $V_{UC}$ = -0.18 V, the dip is so broad because of strong delocalization of the charge over both dots, and the corresponding honeycomb diagram is shown in Fig. 3(b). These results suggest that the size and the width of the dip reflect the strength of the quantum mechanical coupling.

Figure 3(d) schematically shows the variations of the bonding- and antibonding-state energies, $E_B$ and $E_A$, respectively, (upper panel) and the charge displacement of the ground state (middle panel) expected as ε is changed. The charge displacement, which can be measured in the conventional transport measurements, corresponds to the first derivative of $E_B$ with respect to ε, and the slope of the curve is determined by the tunnel coupling energy $t_{tunnel}$ [11]. In contrast, in our measurement, the capacitance dip represents the second derivatives of $E_B$ with respect to ε. Therefore, the depth of the dip indicates the quantum capacitance $C_Q$ given by $C_Q \equiv d^2 E / d\varepsilon^2$ [6] and its width representing the tunnel coupling energy. Figure 3(e) shows the full width at half maximum (FWHM) of the dip as a function of $V_{UC}$. For $V_{UC} > -0.21$ V, the width rapidly increases with $V_{UC}$, indicating the capacitance signal is governed by $C_Q$ and the dip width represents $t_{tunnel}$. The exponential function on $V_{UC}$ is consistent with the previous reports for gate-defined tunneling barriers [8]. For $V_{UC} < -0.21$ V, on the other hand, the width saturates at ~36 μeV. The minimum width may be limited by the large excitation voltages $V_{DQD}$ and $V_{QPC}$ used here (0.1 and 0.56 meV, respectively). The successful measurement of quantum capacitance can be used to measure the tunneling coupling as well as to develop a qubit detector.

In summary, we have studied the tunneling dynamics in a DQD using wide-band capacitance measurements. In the interdot tunnel coupling regime, the quantum capacitance reflecting the strength of quantum-mechanical coupling is measured. This technique can also be used to distinguish the spin as well as charge states of a DQD, since the energy $E$ of the singlet and triplet states of a two-electron system have different dispersion as a function of ε

[12].

This work was supported by the SCOPE from the Ministry of Internal Affairs and Communications of Japan.

[References]

[Figure captions]

[Figure 1]

(a) Schematic illustration of the experimental setup. The measurements were performed at zero external magnetic field with the device cooled in a dilution refrigerator to an electron temperature of 130 mK. (b) Energy diagrams of the dot and the lead when the applied $V_{DQD}$ is high (upper) and low (lower). (c) Schematic of shape of $V_{DQD}$ (upper) and the response of $\Delta Q_{QDR}$ (lower). $T_{rep}$ and $\Gamma$ is the repetition time and the tunneling rate, respectively. (d) $I_{QPC}$ as a function of $V_{UR}$ with $V_{sd\text{-}DQD} = 0.5$ mV and $V_{sd\text{-}QPC} = 0.7$ mV. $N_R$ indicates electron number in the right dot. $V_{UC}$ and $V_{UL}$ are fixed at -0.25 and -0.2 V, respectively. (e) $<I_{QPC}>$ as a function of $V_{UR}$ at $f_{op} = 1$ kHz. The amplitudes of pulse voltages $V_{DQD}$ and $V_{QPC}$ are 0.34 and 0.56 mV, respectively. The peak marked with an asterisk comes from another local potential minimum.

[Figure 2]

(a) $<I_{QPC}>$ as a function of $V_{UR}$ at several $f_{op}$ with $V_{DQD}=0.44$ mV and $V_{QPC}=0.56$ mV. The peak marked with an asterisk comes from another local potential minimum. (b) $\delta I_{peak}$ as a function of $f_{op}$. The solid lines indicate numerically calculated ones, and the dashed line is a guide for the eyes. (c) $\delta I_{CB}$ as a function of $\theta$ at several $f_{op}$ with $V_{DQD}=0.44$ mV and $V_{QPC}=0.56$ mV. (d) GS of $\delta I_{peak}$ as a function of $\theta$ for peak I. The arrows show the minimum points.

[Figure 3]

(a) $<I_{QPC}>$ as a function of $V_{UL}$ and $V_{UR}$ at $V_{UC}=-0.2$ V, $f_{op} = 1$ kHz, $V_{DQD} = 0.1$ mV, and $V_{QPC}=0.56$ mV. Lines A to C indicate the border of the honeycomb structure. The region circled by the dashed line is described in the text. (b) Similar $<I_{QPC}>$ as a function of $V_{UL}$ and $V_{UR}$ at $V_{UC}$=-0.18 V. (c) $\langle I_{QPC} \rangle$ as a function of the bias offset ε at several $V_{UC}$. (d) Schematic energy diagram as a function of the bias offset ε. $E_A$, $E_B$, $E_R$, and $E_L$ correspond to the energy states of the antibonding, bonding, right-dot, and left-dot, respectively. The direction of the gate sweep ε is defined as shown in Fig. 3(a). (e) The full width at half maximum (FWHM) of the dip as a function of $V_{UC}$. The dashed line indicates the fitted linear line.

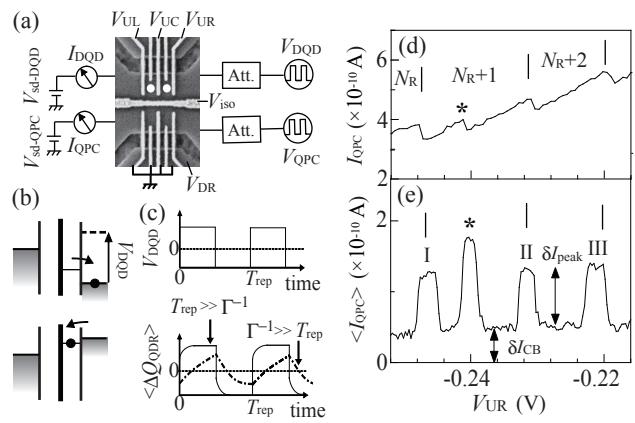

Figure 1

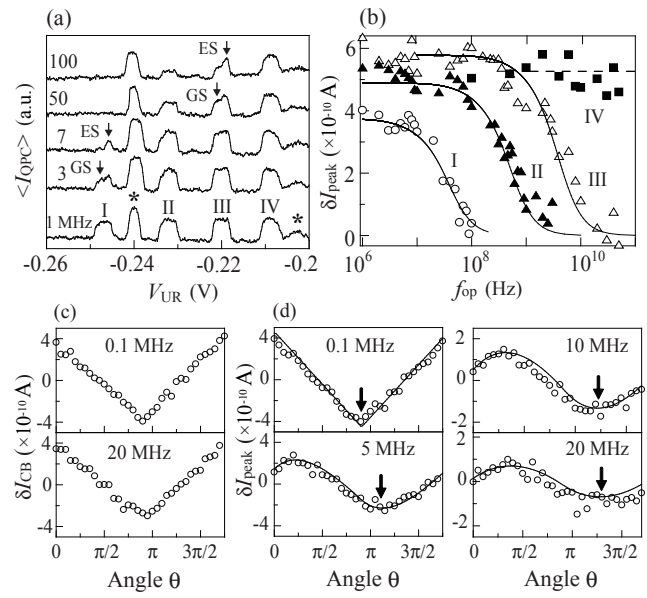

Figure 2

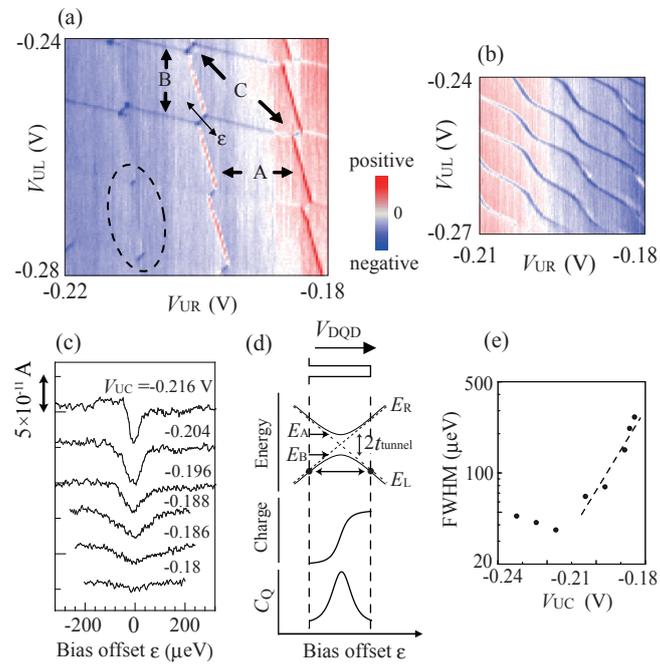

Figure 3